\documentclass[11pt,a4paper]{article}
\usepackage{jcappub}

\usepackage{amssymb,amsmath,graphicx}
\usepackage{bm}
\usepackage{epsfig}

\def\beq{\begin{equation}}
\def\eeq{\end{equation}}
\def\ber{\begin{eqnarray}}
\def\eer{\end{eqnarray}}
\def\benu{\begin{enumerate}}
\def\eenu{\end{enumerate}}
\def\nn{\nonumber}
\def\l{\left}
\def\r{\right}
\def\d{{\rm d}}

\def\pa{\partial}
\def\f{\frac}
\def\mpl{M_{p}}

\def \lleq {\lower0.9ex\hbox{ $\buildrel < \over \sim$} ~}
\def \ggeq {\lower0.9ex\hbox{ $\buildrel > \over \sim$} ~}

\def\prd{{Phys.\@ Rev.\@ D\ }}

\def\ie {{\it ie}}

\title{Consistency relation in power law G-inflation}

\author{Sanil Unnikrishnan\footnote{Current address: Department of Physics, The LNM Institute of Information Technology, Jaipur-302031, India. Email: sanil@lnmiit.ac.in}}
\author{and S.~Shankaranarayanan}
\affiliation{School of Physics, Indian Institute of Science Education and Research,
Thiruvananthapuram 695016, India}
\emailAdd{sanil@iisertvm.ac.in}
\emailAdd{shanki@iisertvm.ac.in}

\date{\today}

\abstract{
In the standard inflationary scenario based on a minimally coupled scalar field, canonical or non-canonical, the subluminal propagation of speed of scalar perturbations ensures the following consistency relation: $r \leq - 8n_{_T}$, where $r$ is the tensor-to-scalar-ratio and $n_{_T}$ is the spectral index for tensor perturbations. However, recently, it has been  demonstrated  that this consistency relation could be violated in Galilean inflation models even in the absence of superluminal propagation of  scalar perturbations. It is therefore interesting to investigate whether the subluminal propagation of scalar field perturbations impose any bound on the ratio $r/|n_{_T}|$ in G-inflation models. In this paper,  we derive the consistency relation for a class of G-inflation models that lead to power law inflation. Within these class of models, it turns out that one can have $r > - 8n_{_T}$ or $r \leq - 8n_{_T}$ depending on the model parameters. However, the subluminal propagation of speed of scalar field perturbations, as required by causality,  restricts  $r \leq -(32/3)\,n_{_T}$.
}

\arxivnumber{1311.0177}

\begin{document}
\maketitle

\section{Introduction}\label{sec:intro}

The inflationary paradigm not only heals the Big Bang theory afflicted with the horizon, flatness and monopole problems~\cite{models,other models}, its prediction of a nearly scale invariant cosmological perturbations~\cite{pert} has remarkably been verified by numerous cosmological observations with  the recent one being the Cosmic Microwave Background (CMB) observation from the \emph{Planck} mission~\cite{Planck-inflation,Planck-cosmo-parameters}.
In spite of this, one is yet to identify the primary source of matter field that caused inflation, although numerous viable models have been proposed~\cite{Martin-2013,Anupam-2011,Riotto-2002,Martin-2004,Bassett-2005,Kinney-2009,Sriram-2009,Baumann-2009}.
In most of these proposed models, inflation is driven by a minimally coupled scalar field.
Broadly speaking, all minimally coupled single scalar field models of inflation can be divided into the following three classes:
\begin{enumerate}
  \item[$(i)$] Canonical scalar field models whose Lagrangian is of the form ${\cal L}\,=\, (1/2)\pa_{\mu}\phi\, \pa^{\mu}\phi\, -\, V(\phi)$.
  \item[$(ii)$] Non-canonical scalar field models~\cite{Garriga-1999,Picon-1999,sanil-2012} in which the Lagrangian ${\cal L}\,=\,{\cal L}(\phi, X)$  is a generic function of the field $\phi$ and the kinetic term $X\,=\, (1/2)\pa_{\mu}\phi\,\pa^{\mu}\phi$.
  \item[$(iii)$] Galilean models of inflation or G-inflation, also known as kinetic gravity braiding models~\cite{Deffayet-2010,G-inf 1st paper}, described by the Lagrangian ${\cal L}\,=\,K(X,\phi)\,+\,G(X,\phi)\Box\phi$, where $K(X,\phi)$ and $G(X,\phi)$ can be an arbitrary function of $\phi$ and the kinetic term $X$.
\end{enumerate}

In fact, the third case above is the most general class of  models describing inflation and  contains the other two cases.
It should be noted that unlike the first two cases, the Lagrangian of the Galilean field contains the second order derivative of the field $\phi$.
However, it turns out that the resultant equation of motion for the scalar field  still remains at the second order as the higher order derivative terms cancel away.
It is one of the possible scalar field models in curved space time that contains higher order terms in the Lagrangian, but still maintains a second order equation for both metric and the field~\cite{Deffayet-2009b}, similar to the Gauss-Bonnet term in the gravity action~\cite{Lovelock-1971}.

It should be noted that a Galilean field corresponds to those class of scalar field models which  are invariant in a Minkowski space time under the Galilean type field transformation, \emph{viz}.~$\phi\, \rightarrow\, \phi\, +\, b_{\mu}x^{\mu}\, +\, c$, where $c$ is a constant and $b_{\mu}$ is a constant vector~\cite{Nicolis-2009,Deffayet-2009a}. Note that the transformation, $~\phi\, \rightarrow \phi\, +\, b_{\mu}x^{\mu}\, + \,c~$, corresponds to shifting the field derivative by a constant vector $b_{\mu}$ similar to the standard  Galilean transformation $\dot{\vec{x}} \rightarrow \dot{\vec{x}} + \vec{V}$ in particle mechanics.
One of the type of scalar field model admitting this type of invariance has Lagrangian of the form ${\cal L}_{3} \propto  X\Box\phi$~\cite{Tsujikawa-2010}.
The Lagrangian of the G-inflation field contains the term $G(X,\phi)\Box\phi$ which can be viewed as the generalization of this type of Galilean interaction, although a generic $G(X,\phi)$ may not admit invariance under $~\partial_{\mu}\phi \rightarrow \partial_{\mu}\phi + b_{\mu}$. Nevertheless, these scalar fields are dubbed as Galilean fields.

The three class of models of inflation discussed above can be characterized by observables such as $n_{_{S}}$, $n_{_{T}}$  and $r$, where $\l\{n_{_{S}},~n_{_{T}}\r\}$ are the spectral indices for scalar and tensor perturbations, respectively, and $r$ is the tensor-to-scalar ratio~\cite{Starobinsky-1979,Starobinsky-1985}.
For the first two classes of inflation models discussed above, namely, the canonical and the non-canonical scalar field models, it turns out that $r$ and $n_{_{T}}$ satisfy the following consistency relation $r \leq - 8n_{_T}$~\cite{Sasaki-1995}. This is,  in fact, the consequence of the subluminal propagation\footnote{It is also possible to violate the consistency relation $r \leq - 8n_{_T}$ in standard canonical inflation models, even in the absence of superluminal propagations, if additional fields generates perturbations~\cite{Kobayashi-2013}.} of the scalar field perturbations. In models which lead to superluminal speed of sound\footnote{The question of whether or not superluminal propagation of scalar field perturbations violates causality is debated in the literature~\cite{superluminal-controversies}.}, one gets $r > - 8n_{_T}$, see for instance~\cite{Mukhanov-2006}.
However, it is recently demonstrated that for the G-inflation models~\cite{G-inf 1st paper,Kamada-2010}, this consistency relation can be violated even after ensuring that the speed of sound of scalar field perturbations is  subluminal.
It is therefore interesting to investigate whether the subluminal propagation of scalar field perturbations in G-inflation models put any upper limit on the ratio $r/(8|n_{_{T}}|)$. For the case of canonical and non-canonical scalar field models of inflation, it is known that the upper limit of the above mentioned ratio is unity. However, it is  not known whether such an upper limit exists for G-inflation models.

To address this issue, we consider a restricted class of analytically solvable G-inflation model in which the $K(X,\phi)$ term only contributes the potential $V(\phi)$ and $G(X,\phi)\,\propto\, X^{n}$ and it is independent of the field $\phi$. This will ensure that the contribution of the k-inflation term $K(X,\phi)$ is minimal and any violation of the consistency relation can therefore be attributed to the G-inflation term $G(X,\phi)$. For this class of model, we obtain the form of the potential $V(\phi)$ which can lead to power law inflation. For power law solution, it is possible to arrive at an exact inflationary consistency relation between $r$ and  $n_{_{T}}$ without imposing the slow roll condition.
We therefore derive such a consistency relation for the class of G-inflation model driving power law inflation.
The limit on the ratio $r/(8|n_{_{T}}|)$ can thus be found by imposing the  restriction that the speed of sound for scalar field perturbations is subluminal.

This paper is organised as follows.
In Sec.~\ref{sec: G-inf prelm}, all the basic equations describing the field dynamics of a generic G-inflation model in a spatially flat universe is discussed. In Sec.~\ref{sec: GPLI}, we introduce a specific class of G-inflation model which can drive power law inflation.
The inflationary consistency relation for such a power law model is derived Sec.~\ref{sec: CR in GPLI}.
In Sec.~\ref{sec: GPLI vs NCPLI}, the observationally viable Galilean inflationary scenario is compared with those based on non-canonical scalar field settings.
Finally, the main results of this paper are highlighted in Sec.~\ref{sec: conclusions}.
The derivation of the expression for the speed of sound in G-inflation models is described in  Appendix~\ref{sec: speed of sound}.
Throughout this paper, we shall adopt the metric signature of $\l(+\,-\,-\,-\,\r)$ and we  express every equations in natural units thereby setting $\hbar\,=\,c\,=\,1$. In such units the reduced Planck mass $\mpl$ is defined as $\mpl\,=\, \l(8 \pi G\r)^{-1/2}$.

\section{G-Inflation Preliminaries}
\label{sec: G-inf prelm}
We  consider the following Einstein-Hilbert action with a Galilean scalar field:
\beq
{\cal S}\,=\,-\frac{\mpl^{2}}{2}\int\!R\,\sqrt{-g}\;\d^{4}x\,+\, \int\!\sqrt{-g}\;\d^{4}x\;{\cal L}(\phi,X,\Box\phi)\,,
\label{eqn: action}
\eeq
where
\beq
{\cal L}(\phi,\,X,\,\Box\phi)\,=\, K(X,\phi)\,+\, G(X,\phi)\Box\phi~,
\label{eqn: G-Lagrangian}
\eeq
is the Lagrangian density of the G-inflation field~\cite{G-inf 1st paper,Deffayet-2010}.
In the above Lagrangian the function $K(X,\phi)$ and $G(\phi , X)$ can, in general,  be an arbitrary function of the field $\phi$ and the kinetic term $X= (1/2)\pa_{\mu}\phi\; \pa^{\mu}\phi$. The form of the Lagrangian (\ref{eqn: G-Lagrangian}) takes care of almost all minimally coupled scalar field models of inflation. When $G(X,\phi)\,=\,0$, the model represent non-canonical scalar field inflation also known as k-inflation\footnote{The class of models with ${\cal L}\,=\,{\cal L}(\phi, X)$ are  known as k-inflation or kinetic inflation  since in some of these models, first introduced in Ref.~\cite{Garriga-1999,Picon-1999}, it is the kinetic term in the Lagrangian which drives inflation. Therefore, in the k-inflation models described in Ref.~\cite{Garriga-1999,Picon-1999}, ${\cal L}(X,\phi)\,\rightarrow\,0$  as $X\,\rightarrow\,0$. However, not all non-canonical models satisfies this criteria, see for instance Refs.~\cite{DBI,Steer-2004,sanil-2012}. Nevertheless, a generic model with ${\cal L} = {\cal L}(X,\phi)$, may be referred as either k-inflation or as non-canonical model of inflation.}~\cite{Garriga-1999,Picon-1999} and in addition if $K(X,\phi)\,=\, X - V(\phi)$ it reduces to the standard canonical scalar field model of inflation.

The Lagrangian (\ref{eqn: G-Lagrangian}) contains the second  derivative of the field $\phi$. After removing the boundary term $\l(G\,\pa^{\mu}\phi\r)_{;\,\mu}$ one gets the following equivalent Lagrangian density~\cite{Deffayet-2010}:
\beq
{\cal L}^{(\mathrm{e})}\,=\,K - 2XG_{\phi} - G_{X}\partial^{\mu}X\,\partial_{\mu}\phi~,
\label{eqn: eqvlnt G-Lagrangian}
\eeq
where the notations such as $G_{\phi}$ denotes $\partial G/\partial \phi$.
Note that when $G_{X}\,=\,0$, the above Lagrangian is a function only of $X$ and $\phi$,  and hence in this scenario it is equivalent to a k-inflation model.
However, when $G_{X}\,\neq\,0$, the Lagrangian (\ref{eqn: G-Lagrangian}) contains the second order derivative in $\phi$  after removing the boundary term and hence, in this case the model is phenomenologically distinct from the k-inflation models. We will be considering such a case where $G_{X}\,\neq\,0$.

From the action~(\ref{eqn: action}), the field equation for $\phi$ is given by
\beq
\frac{\partial}{\partial\phi}\l({\cal L}\sqrt{-g}\r)\,-\,\partial_\mu\l(\frac{\partial{\cal L}\sqrt{-g}}{\partial(\partial_\mu \phi)}\r)\,+\, \partial_{\mu}\partial_{\nu}\l(\frac{\partial{\cal L}\sqrt{-g}}{\partial(\partial_{\mu}\partial_{\nu} \phi)}\r)\,=\,0~. \label{eqn: Lagn-eqn}
\eeq
On substituting the Lagrangian density (\ref{eqn: G-Lagrangian}) in the above equation, we get
\ber
&&\l(2G_{\phi} -2XG_{X\phi} - K_{X}\r)\Box\phi\,+\,
\l(2G_{X\phi} -K_{XX}\r)\partial^{\mu}\phi\partial_{\mu} X \,+\,
2X\l(G_{\phi\phi}-K_{X\phi}\r)\, +\, K_{\phi}\,+\,\hspace{1cm}\nn\\
&&~~G_{X}\l[(\partial^{\mu}\phi)_{;\nu}(\partial^{\nu}\phi)_{;\mu}- (\Box\phi)^{2} + R_{\mu\nu}\partial^{\mu}\phi\partial^{\nu}\phi\r]\,+\,
G_{XX}\l[\partial^{\mu}X\partial_{\mu}X - (\partial^{\mu}\phi\partial_{\mu} X)\Box\phi \r] = 0
\label{eqn: G_I EOM}
\eer
Note that one gets the same equation of motion if instead of ${\cal L}$ from Eq.~(\ref{eqn: G-Lagrangian}) one substitutes the equivalent Lagrangian (\ref{eqn: eqvlnt G-Lagrangian}) in Eq.~(\ref{eqn: Lagn-eqn}). It is also important to note that although we started with the action~(\ref{eqn: action}) in which the field $\phi$ is minimally coupled to gravity, the resulting field equation contains a term $R_{\mu\nu}\partial^{\mu}\phi\,\partial^{\nu}\phi$ indicating a coupling between the Ricci tensor and the kinetic term. It is for this reason, these class of models are also known as kinetic gravity braiding models~\cite{Deffayet-2010}.

On varying the action~(\ref{eqn: action}) with respect to the metric $g_{\mu\nu}$ gives the Einstein's equation $G_{\mu\nu}=(8\pi G)\,T_{\mu\nu}$, where the energy momentum tensor $T_{\mu\nu}$ is defined as
\beq
T_{\mu\nu}\,=\,  \frac{2}{\sqrt{-g}}\l[\frac{\partial}{\partial g^{\mu\nu}}\l({\cal L}\sqrt{-g}\r)\;-\;
\partial_{\alpha}\l(\frac{\partial{\cal L}\sqrt{-g}}{\partial (g^{\mu\nu}_{~~,\alpha})}\r)
\r].
\label{eqn: EM tensor def}
\eeq
Substituting ${\cal L}$ from Eq.~(\ref{eqn: G-Lagrangian}) or ${\cal L}^{(\mathrm{e})}$ from Eq.~(\ref{eqn: eqvlnt G-Lagrangian}) in the above equation gives
\ber
T^{\mu}_{~~\nu}\,=\, -\,G_{X}\l[\partial^{\mu}\phi\,\partial_{\nu} X + \partial^{\mu}X\,\partial_{\nu}\phi\r]\,-\,
\l[2G_{\phi} - G_{X}\square\phi - K_{X}\r] \partial^{\mu}\phi\,\partial_{\nu}\phi\nn\\
\,+\;\l[-K + G_{X}\partial^{\alpha}X\,\partial_{\alpha}\phi + 2XG_{\phi}\r]\delta^{\mu}_{~~\nu}~.
~~~~~~~~~~~~~~~~~~~~~~~~~~~~~~~~
\label{eqn: EM tensor G-field}
\eer
Considering a spatially flat Friedmann-Robertson-Walker (FRW) universe described by the line element
\beq
\d s^2 = \d t^2-a^{2}(t)\; \l[\d x^2 + \d y^2 + \d z^2\r],
\label{eqn: FRW}
\eeq
the above expression (\ref{eqn: EM tensor G-field}) for the energy momentum tensor takes the diagonal form:
\beq
T^{\mu}_{~~\nu}\,=\, \mathrm{dia}\l\{\rho_{_\phi}, - p_{_\phi},-  p_{_\phi}, - p_{_\phi}\r\},
\eeq
where the energy density $\rho_{_\phi}$ and the pressure $p_{_\phi}$ are given by
\ber
\rho_{_\phi}\,&=&\,2XK_{X} - K - 2XG_{\phi} + 6H\dot{\phi}XG_{X}~,
\label{eqn: rho G}
\\
p_{_\phi}\,&=&\,K - 2XG_{\phi} - 2X\ddot{\phi}G_{X}~.
\label{eqn: p G}
\eer
In the above two equations, $H\,\equiv\, \dot{a}/a$, where $a(t)$ is the scale factor.
The Einstein's equation $G_{\mu\nu}=(8\pi G)\,T_{\mu\nu}$ implies that the scale factor $a(t)$ satisfies the following Friedmann equations:
\ber
\l(\frac{\dot{a}}{a}\r)^{2} &=& \l(\frac{8 \pi G}{3}\r)\rho_{_{\phi}},\label{eqn: Friedmann eqn1}\\
\frac{\ddot{a}}{a} &=& -\l(\frac{4 \pi G}{3}\r)\l(\rho_{_{\phi}} + 3\,p_{_{\phi}}\r)\,,\label{eqn: Friedmann eqn2}
\eer
Note that the expression for the energy density $\rho_{_\phi}$, as described in Eq.~(\ref{eqn: rho G}), contains $H$. Therefore, the first Friedmann equation~(\ref{eqn: Friedmann eqn1}) describes a quadratic equation for $H$ unlike the usual case when one considers canonical or non-canonical scalar field models.
To ensure that $H$ is real and positive definite, it is necessary that the following conditions are satisfied:
\beq
\frac{\dot{\phi}\,X\,G_{X}}{\mpl^{2}}\,\geq\,0
\eeq
\beq
\l(\frac{\dot{\phi}\,X\,G_{X}}{\mpl^{2}}\r)^{2} \,\geq\, \l(\frac{1}{3\mpl^{2}}\r)\l(K - 2XK_{X} + 2XG_{\phi}\r)
\eeq
With the above condition, the  Friedmann equation~(\ref{eqn: Friedmann eqn1}) becomes
\beq
H\,=\,\frac{\dot{\phi}\,X\,G_{X}}{\mpl^{2}} + \l(\frac{1}{\mpl^{2}}\r)
\l[\l(\dot{\phi}\,X\,G_{X}\r)^{2} + \l(\frac{\mpl^{2}}{3}\r)\l(2XK_{X} -K - 2XG_{\phi} \r)\r]^{1/2}
\label{eqn: H}
\eeq

Moving on to the field equation for $\phi$ as described in Eq.~(\ref{eqn: G_I EOM}), notice that because of the term $R_{\mu\nu}\partial^{\mu}\phi\partial^{\nu}\phi$, the equation of motion for $\phi$ contains terms proportional to $\dot{H}$. This can be eliminated using the two Friedmann equations~(\ref{eqn: Friedmann eqn1}) and (\ref{eqn: Friedmann eqn2}). The equation of motion for $\phi$ can then be expressed as
\beq
C_{_1}\ddot{\phi}\, +\, C_{_2}(3H\dot{\phi})\, +\, 2XC_{_3}\, -\, K_{\phi}\,=\,0
\label{eqn: KgEqn for G}
\eeq
where
\ber
C_{_1}\,&=&\, K_{X} + 2XK_{XX} - 2G_{\phi} -  2XG_{X\phi} + 6H\dot{\phi}\l(G_{X} + XG_{XX}\r)  + \frac{6X^{2}G_{X}^{2}}{\mpl^{2}}~,\nn\\
C_{_2}\,&=&\, K_{X}  - 2G_{\phi} + 2XG_{X\phi} + 3H\dot{\phi}G_{X} - \frac{6X^{2}G_{X}^{2}}{\mpl^{2}}~,\\
C_{_3}\,&=&\, K_{X\phi} - G_{\phi\phi} - \l(\frac{3XG_{X}}{\mpl^{2}}\r)\l(K_{X} - 2G_{\phi}\r)~.\nn
\eer
Eqs.~(\ref{eqn: H}) and (\ref{eqn: KgEqn for G}) forms the two closed set of equations describing the evolution of $a(t)$ and $\phi(t)$.
For the case of canonical scalar field which corresponds to choosing $K(X,\phi)\,=\,X\,-\,V(\phi)$ and $G(X,\phi)\,=\,0$, the field equation~(\ref{eqn: KgEqn for G}) reduces to the standard Klein-Gordon equation {\it viz.\ }$\ddot{\phi} + 3H\dot{\phi}  + V_{\phi}\,=\,0$.

\section{Power law G-Inflation}
\label{sec: GPLI}
Our aim in this paper is to derive an exact consistency relation in G-inflation models without assuming slow roll. It is possible to do so in the case of power law inflation for which one can obtain an exact analytical expression for $r$ and $n_{_T}$. For the case of kinetic power law inflation, such an exact consistency relation  is derived in Ref.~\cite{Garriga-1999}. It follows from the analysis of Ref.~\cite{Garriga-1999} that $r \leq - 8n_{_T}$ when the speed of sound is subluminal.

Although, G-inflation corresponds to a wider class of models with generic $K(X,\phi)$ and $G(X,\phi)$ in the Lagrangian~(\ref{eqn: G-Lagrangian}),
to understand the exact reason for the violation of the consistency relation as noted in Ref.~\cite{Kamada-2010}, it is necessary to minimize the contribution from  the k-inflation term  $K(X,\phi)$. If we eliminate the contribution of the $K(X,\phi)$ altogether, and chose $G(X,\phi)\,\propto\, X^{n}$, then it lead to $c_{_s}^{2} < 0$ thereby making the system violently unstable. This also happens when $K(X,\phi) = 0$ and $G(X,\phi) = g(\phi)X^{n}$ with an integer value for $n$.
For this reason we restrict ourself to a subclass of these models with $K(X,\phi) = - V(\phi)$ and $G(X,\phi)\,\propto\, X^{n}$. The Lagrangian of this restricted class of G-inflation model considered in this paper is therefore given by:
\beq
{\cal L}(\phi,\,X,\,\Box\phi)\,=\,\frac{X^{n}\,\Box\phi}{M^{4n-1}}\; -\; V(\phi)~,
\label{eqn: Lagrangian GPLI}
\eeq
where $n$ and $M$ are parameters of the model with $n$ being dimensionless and $M$ has dimensions of mass.
We will now obtain the form of the potential $V(\phi)$ which can drive power law inflation wherein the scale factor evolves as
\beq
a(t)\, \propto\, t^q~; ~~~~~~ q\,>\,1\nn
\eeq

For model~(\ref{eqn: Lagrangian GPLI}), the energy density $\rho_{_\phi}$ and the  pressure $p_{_\phi}$ turns out to be:
\ber
\rho_{_\phi}\,&=&\,\l(\frac{3\, n\, H}{2^{n - 1}}\r)\l(\frac{\dot{\phi}^{2n + 1}}{M^{4n - 1}}\r)
\,+\, V(\phi),
\label{eqn: rho GPLI}
\\
p_{_\phi}\,&=&\,- \l(\frac{n\, \ddot{\phi}}{2^{n-1}}\r)\l(\frac{\dot{\phi}^{2n}}{M^{4n - 1}}\r)\,-\, V(\phi).
\label{eqn: p GPLI}
\eer
When $a(t)\, \propto\, t^q$, the Friedmann equations~(\ref{eqn: Friedmann eqn1}) and (\ref{eqn: Friedmann eqn2}) implies that
\ber
\rho_{_{\phi}} &=& \frac{3\,\mpl^2\,q^{2}}{t^{2}}~,\label{eqn: rho-t GPLI}\\
p_{_{\phi}} &=& w\,\rho_{_{\phi}}~, ~~
\label{eqn: p-t GPLI}
\eer
where $w\, =\, 2/(3q) - 1$. From Eqs.~(\ref{eqn: rho GPLI}) to (\ref{eqn: p-t GPLI}), it follows that
\beq
\frac{\mpl^{2}}{t}\,=\,\l(\frac{3\, n\,\dot{\phi}^{2n + 1}}{2^{n}\,M^{4n - 1}}\r)\l(1 - \frac{\ddot{\phi}}{3H\dot{\phi}}\r)~,
\label{eqn: phi dot eqn}
\eeq
\beq
V(\phi)\,=\,-\l(\frac{n\,2^{2 - n}}{1 + w}\r)\l(\frac{\dot{\phi}^{2n + 1}}{M^{4n - 1}}\r)
\l(\frac{1}{t}\r)\,+\, \frac{4\,\mpl^{2}~~}{3\,(1+w)^{2}\,t^{2}}~.
\label{eqn: V phi eqn}
\eeq
It can be verified that Eq.~(\ref{eqn: phi dot eqn}) admit solution of the form:
\beq
\phi(t)\,=\, A\,\mpl^{\beta + 1}\,t^{\beta}~;~~~~\mathrm{where}~~~\beta\,=\, \frac{2n}{2n + 1}~,
\label{eqn: phi soln}
\eeq
and
\beq
 A\,=\, \l(\frac{2n + 1}{2n}\r)
\l\{
\l(\frac{2^{n}}{3\,n}\r)\l(\frac{2\,(2n + 1)}{2\,(2n + 1) + (1 + w)}\r)\l(\frac{M}{\mpl}\r)^{4n -1}
\r\}^{\frac{1}{2n + 1}}~.
\label{eqn: A}
\eeq
Substituting the solution~(\ref{eqn: phi soln}) in Eq.~(\ref{eqn: V phi eqn}), we get the following form of the potential:
\beq
V(\phi)  = \frac{V_{_0}}{(\phi/\mpl)^{s}}~;~~~~\mathrm{where}~~~s  = \frac{2\,n + 1}{n}~,
\label{eqn: V-phi GPLI}
\eeq
and
\beq
V_{_0}\,  =\, \l(\frac{4\,A^{s}\,\mpl^{4}}{3\,(1 + w)^{2}}\r)
\l(\frac{1 - (4n+1)\,w}{2\,(2n + 1) + (1 + w)}\r)~.
\label{eqn: V-0 GPLI}
\eeq
In the model~(\ref{eqn: Lagrangian GPLI}) with the above form of the potential, although tedious, it is straight forward to verify that the solution~(\ref{eqn: phi soln}) with $a(t)\, \propto\, t^q$ satisfy both the Friedmann's equations (\ref{eqn: Friedmann eqn1}), (\ref{eqn: Friedmann eqn2}) and the scalar field equation~(\ref{eqn: KgEqn for G}).
Hence, in the Lagrangian~(\ref{eqn: Lagrangian GPLI}), an inverse power law potential of the form~(\ref{eqn: V-phi GPLI}) can drive power law inflation with $a(t)\, \propto\, t^q$.
It is interesting to note that  such inverse power law potentials also drive power law inflation in non-canonical scalar field models, see for instance power law models described in Refs.~\cite{Garriga-1999,Picon-1999,sanil-2013}.
However, in the case  canonical scalar field driven inflation, an inverse power law potential leads to intermediate inflation~\cite{barrow90,muslimov90}.

\section{Consistency relation in Galilean Power law inflation}
\label{sec: CR in GPLI}
In this section we shall derive the consistency relation for the Galilean power law inflation model described in the preceding section.
However, before moving on the specific model~(\ref{eqn: Lagrangian GPLI}), let us first consider the generic G-inflation scenario.
To obtain the scalar and tensor perturbations generated by the inflation field, we consider the following FRW line element with metric perturbations~\cite{Bardeen-1980,Kodama-1984,Mukhanov-1992}
\beq
\d s^2
= (1+2\, A)\,\d t ^2 - 2\, a(t)\, (\pa_{i} B )\; \d t\; \d x^i\,
-a^{2}(t)\; \l[(1-2\, \psi)\; \delta _{ij}+ 2\, \l(\pa_{i}\, \pa_{j}E \r) + h_{ij}\r]\,
\d x^i\, \d x^j\nn
\eeq
where $A$, $B$, $\psi$ and $E$ are scalar degree of metric perturbation and $h_{ij}$ is the tensor perturbations.
The vector perturbations are ignored as it is known that scalar fields do not lead to such perturbations.
The perturbation in the scalar field is defined as
\beq
\phi(\vec{x},t)\, = \,^{(0)}\phi(t)\,+\,\delta\phi(\vec{x},t)~\,
\eeq
where $^{(0)}\phi(t)$ is the background field which, for the G-inflation case, satisfies Eq.~(\ref{eqn: KgEqn for G}).
The perturbation $\delta\phi(\vec{x},t)$ being a  gauge dependent quantity, one generally introduce  the following gauge invariant quantity $\mathcal{R}$ known as curvature perturbation:
\beq
\mathcal{R} \equiv \psi + \l(\frac{H}{\dot{\phi}}\r)\delta \phi~.
\eeq
In the generic model with the Lagrangian~(\ref{eqn: G-Lagrangian}), the second order action for the curvature perturbation $\mathcal{R}$ turns out to be~\cite{G-inf 1st paper,Deffayet-2010,Brandenberger-2012}
\beq
{\cal S}^{(2)}\,=\,\frac{1}{2}\int\,\d^{3}x\,\d \eta\,z^{2}\l[\,\mathcal{R}'^{2} - c_{_s}^{2}(\partial_{i}\mathcal{R})^{2}\,\r],
\label{eqn: action S2}
\eeq
where $\eta$ is the conformal time and  $c_{_s}$ is the speed of sound for the G-inflation field whose square is given by
\beq
c_{_s}^{2} \equiv \frac{K_{X} - 2G_{\phi}  + 2XG_{X\phi} + 2\ddot{\phi}\l(G_{X} + X G_{XX}\r) + 4H\dot{\phi}G_{X} - 2X^{2}G_{X}^{2}/\mpl^{2}}
{K_{X} + 2XK_{XX} -  2G_{\phi}  - 2XG_{X\phi} + 6H\dot{\phi}\l(G_{X} + X G_{XX}\r) + 6X^{2}G_{X}^{2}/\mpl^{2}}~.
\label{eqn: sound speed G-inf def}
\eeq
In the action~(\ref{eqn: action S2}), the function $z$ is defined as
\beq
z\, \equiv\, \l(\frac{a\,\dot{\phi}\,\sqrt{\mathcal{F}}}{c_{_s}\,H}\r)
\l[1 - \l(\frac{\dot{\phi}^{3}\,G_{X}}{2\,H\,\mpl^{2}}\r)\r]^{-1}~,
\label{eqn: z G-inf def}
\eeq
where
\beq
\mathcal{F}\, =\, K_{X} - 2G_{\phi}  + 2XG_{X\phi} + 2\ddot{\phi}\l(G_{X} + X G_{XX}\r) + 4H\dot{\phi}G_{X} - \frac{2X^{2}G_{X}^{2}}{\mpl^{2}}~.
\label{eqn: F G-inf def}
\eeq

From the action~(\ref{eqn: action S2}), one gets the following equation of motion for the curvature perturbation
\beq
\mathcal{R}_{_k}'' + 2\l(\frac{z'}{z}\r)\mathcal{R}_{_k}'  + c_{_s}^{2}\,k^{2}\,\mathcal{R}_{_k} = 0~,
\label{eqn: curvature pert eqn G-inf}
\eeq
where $\mathcal{R}_{_k}$ is the amplitude of the curvature perturbation $\mathcal{R}$  in the Fourier space and $k$ is the wavenumber. Note that just like in the case of canonical or non-canonical scalar field driven inflation, the curvature perturbation $\mathcal{R}$ is also conserved at the superhorizon scales in G-inflation models, for a proof see  Ref.~\cite{sasaki-2011}.

In terms of the  Mukhanov-Sasaki variable $u_{_k} = z\,\mathcal{R}_{_k}$, Eq.~(\ref{eqn: curvature pert eqn G-inf}) becomes:
\beq
u_{_k}'' +  \l(c_{_s}^{2}\,k^{2} -   \frac{z''}{z\;}\r)u_{_k} = 0~.
\label{eqn: MS eqn G-inf}
\eeq
This is exactly identical to the corresponding equation for the k-inflation field~\cite{Garriga-1999}, the difference being $c_{_s}^{2}$ and $z$ in the above equation are different from those that appear in the Mukhanov-Sasaki equation in k-inflation models (see Eq.~(28) in Ref.~\cite{Garriga-1999}).

Similarly, for tensor perturbations one gets the following equation:
\beq
v_{_k}'' +  \l(k^{2} -   \frac{a''}{a}\r)v_{_k} = 0 ~,
\label{eqn: MS eqn tensor k-inf}
\eeq
where $v_{_k} = a h_{_k} (\mpl/2)$, with $h_{_k}$ being the Fourier amplitude of tensor perturbations.
Unlike the case of Mukhanov-Sasaki equation for the scalar variable $u_{_k}$, the above equation is \emph{identically} valid for all minimally coupled scalar field models of inflation since tensor perturbations evolves independent of the scalar perturbations at the linear order.

The  scalar and tensor power spectrum are defined as
\beq
\mathcal{P}_{_{S}}(k) \equiv \l(\frac{k^{3}}{2\pi^{2}}\r)|\mathcal{R}_{_k}|^{2} = \l(\frac{k^{3}}{2\pi^{2}}\r)\l(\frac{|u_{_k}|}{z}\r)^{2} ~,
\label{eqn: scalar PS}
\eeq
\beq
\mathcal{P}_{_{T}}(k) \equiv 2\l(\frac{k^{3}}{2\pi^{2}}\r)|h_{_k}|^{2} = 2 \l(\frac{k^{3}}{2\pi^{2}}\r)\l(\frac{4}{\mpl^{2}}\r)\l(\frac{|v_{_k}|}{a}\r)^{2}.
\label{eqn: tensor PS}
\eeq
Furthermore, one defines the scalar and tensor spectral index as:
\ber
n_{_{S}} - 1 &\equiv&  \frac{\d\, \mathrm{ln} \mathcal{P}_{_{S}}}{\d\, \mathrm{ln} k}~,
\label{eqn: ns definition}\\
n_{_{T}} &\equiv&  \frac{\d\, \mathrm{ln} \mathcal{P}_{_{T}}}{\d\, \mathrm{ln} k}~.
\label{eqn: nt definition}
\eer
And finally the tensor-to-scalar ratio $r$ is defined as
\beq
r \equiv \f{\mathcal{P}_{_{T}}}{\mathcal{P}_{_{S}}}~.
\label{eqn: T-to-S def}
\eeq

It is clear from Eq.~(\ref{eqn: MS eqn G-inf}) that it is the function $z$ which can lead to a different evolution for the mode function $u_{_k}$ in a G-inflation model from those in a k-inflation model which has the same value for the speed of sound for scalar perturbations.
It is therefore illustrative to express the function $z$ defined in Eq.~(\ref{eqn: z G-inf def}) as
\beq
z\,=\, \frac{\tilde{z}}{\lambda}
\label{eqn: z and z tilde}
\eeq
where $\tilde{z}$, defined as
\beq
\tilde{z} \equiv \frac{a\,\l(\rho_{_{\phi}}+ p_{_{\phi}}\r)^{1/2}}{c_{_s}H}~,
\label{eqn: z tilde}
\eeq
is the one that appears in the Mukhanov-Sassaki equation for the k-inflation field~\cite{Garriga-1999} and $\lambda$ is defined as
\beq
\lambda\,=\,
\l[1 - \l(\frac{\dot{\phi}^{3}\,G_{X}}{2\,H\,\mpl^{2}}\r)\r]
\sqrt{\frac{\mathcal{Q}}{2\,X\,\mathcal{F}}}~,
\label{eqn: lambda def}
\eeq
where $\mathcal{F}$ is defined in Eq.~(\ref{eqn: F G-inf def}) and $\mathcal{Q}$ is given by
\beq
\mathcal{Q}\,=\, 2XK_{X}\,-\, 4XG_{\phi} - 2XG_{X}\l(\ddot{\phi} - 3H\dot{\phi}\r).
\label{eqn: Q def}
\eeq

Note that in the case of k-inflation which corresponds to setting $G(X,\phi)\,=\,0$, in the Lagrangian~(\ref{eqn: G-Lagrangian}), one gets $\lambda = 1$. Hence, it is because of this $\lambda$ factor in Eq.~(\ref{eqn: z and z tilde}) that makes the scalar power spectrum in G-inflation different from those in an equivalent k-inflation model which leads to the same background evolution and has the same value for $c_{_s}^{2}$.  Since Eq.~(\ref{eqn: MS eqn tensor k-inf}) is valid for all minimally coupled scalar field models of inflation, the tensor power spectrum in G-inflation is exactly the same as those in an equivalent k-inflation model which leads to the same background evolution. It is for this reason the tensor-to-scalar ratio in G-inflation will be different  from those in an equivalent k-inflation model.
This point will be illustrated in detail for the power law model considered in the preceding section

\begin{figure}[t]
\begin{center}
\scalebox{0.83}[1]{\includegraphics{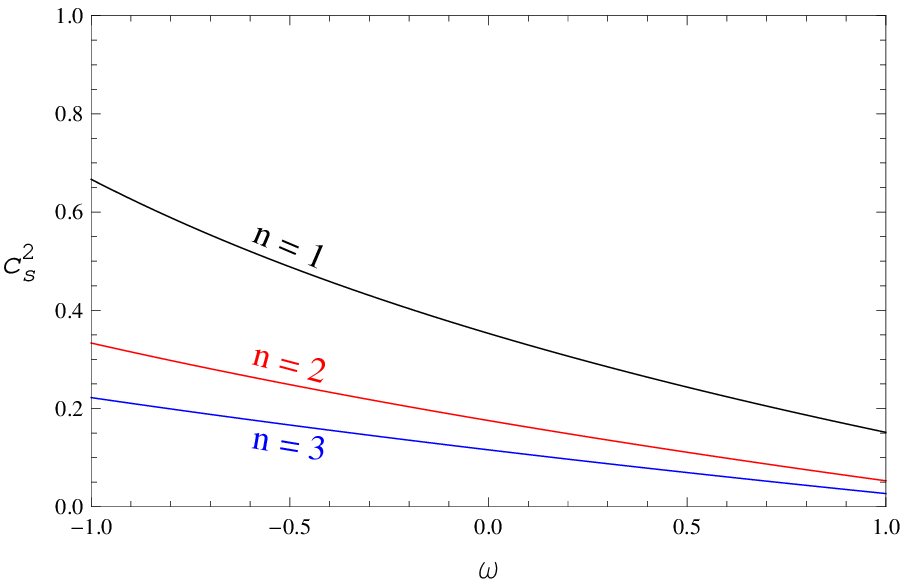}}
\scalebox{0.83}[1]{\includegraphics{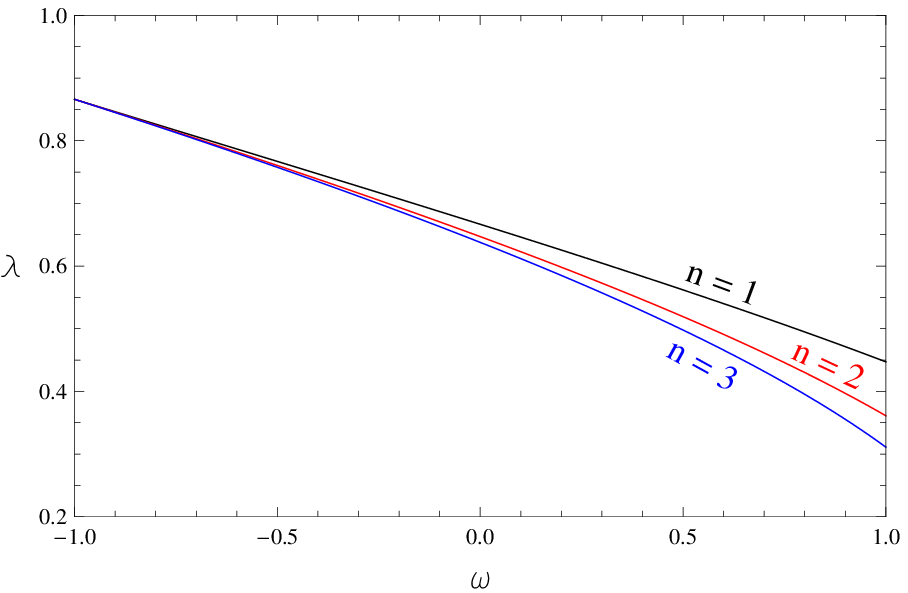}}
\caption{In the left panel,  $c_{_s}^{2}$ for the Galilean power law inflation model, as given in Eq.~(\ref{eqn: sound speed GPLI}), is plotted as a function of equation of state parameter $w$ for different values of $n$ in the Lagrangian~(\ref{eqn: Lagrangian GPLI}). In the right panel $\lambda$ given in Eq.~(\ref{eqn: lambda GPLI}) is plotted as a function of $w$. Note that both $c_{_s}^{2}$ and $\lambda$ are positive definite and they are less than unity.
In this figure, the range of $w$ is from $-1$ to $+1$, although for inflation one only need to consider the regime where $w < -1/3$.}
\label{fig: cs and lambda}
\end{center}
\end{figure}

For the restricted class of G-inflation model~(\ref{eqn: Lagrangian GPLI}) with an inverse power law potential~(\ref{eqn: V-phi GPLI}) driving power law inflation with $a(t)\, \propto\, t^q$, it follows from Eqs.~(\ref{eqn: phi soln}) and (\ref{eqn: sound speed G-inf def}) that
\beq
c_{_s}^{2} = \l(\frac{1}{3(2n + 1)}\r)\l\{
\frac{\l[\,4(2n + 1) - 3n(1 + w)\,\r]\l[\,2(2n + 1) + (1 + w)\,\r] - (2n + 1)^{2}(1 + w)}
{4n(2n + 1) + (1 + w)(4n + 1)}\r\}~,
\label{eqn: sound speed GPLI}
\eeq
where $w$ is the equation of state parameter which is related to parameter $q$ in the power law solution $a(t) \propto t^q$ as $w = -1 \,+\, 2/(3q)$.
It is clear from Eq.~(\ref{eqn: sound speed GPLI}) that the speed of sound is identically constant for the Galilean power law inflation model~(\ref{eqn: Lagrangian GPLI}) and in the slow roll limit which corresponds to $(1 + w)\,\ll\,1$ or equivalently $q\,\gg\,1$, one gets $c_{_s}^{2}\, \simeq\,2/(3n)$. When $n = 1$, the slow roll value of $c_{_s}^{2}$ is $2/3$ and this is consistent with Ref.~\cite{Kamada-2010} which considered Higgs G-inflation. In the left panel of Fig.~\ref{fig: cs and lambda}, $c_{_s}^{2}$ is plotted as a function of $w$.
Note that the speed of sound for the Galilean model is subluminal~\cite{Easson-2013}.
Furthermore, the solution~(\ref{eqn: phi soln}) also implies that the $\lambda$ parameter defined in Eq.~(\ref{eqn: lambda def}) is a constant and is given by
\beq
\lambda\,=\,
\l\{\frac{6\,\l[(2n + 1) - n(1 + w)\r]^{2}}
{8\,(2n + 1)^{2} - (1 + w)\l[16n^{2} + 2n - 3\r] - 3n(1 + w)^{2}}
\r\}^{1/2}~.
\label{eqn: lambda GPLI}
\eeq
The parameter $\lambda$ is plotted as a function of $w$ in the right panel of Fig.~\ref{fig: cs and lambda}.
In the slow roll limit which corresponds to $(1 + w)\,\ll\,1$, the parameter $\lambda\,\rightarrow\,\sqrt{3/4}~$ irrespective of the value of $n$ in the Lagrangian~(\ref{eqn: Lagrangian GPLI}).

For the power law solution $a(t) \propto t^q$, the function $z$ defined in Eq.~(\ref{eqn: z and z tilde}) becomes
\beq
z\,=\, \l(\frac{\mpl\,a}{\lambda\,c_{_s}}\r)\sqrt{\frac{2}{q}}~.
\eeq
With $z$ given by the above expression, the Mukhanov-Sasaki equation~(\ref{eqn: MS eqn G-inf}) can be expressed as
\beq
u_{_k}'' +  \l(c_{_s}^{2}\,k^{2}\; - \;  \frac{\nu^{2} - (1/4)}{\eta^{2}\;}\r)u_{_k} = 0~,
\label{eqn: MS eqn GPLI}
\eeq
where
\beq
\nu\,=\, \frac{3q - 1}{2\,(q - 1)}~.\nn
\eeq
The general solution of the above equation can be expressed as
$$u_{_k} = \sqrt{-c_{_s}k\eta}
\l[\mathrm{C}_{1}\,\mathrm{H}^{^{(1)}}_{\nu}\l(-c_{_s}k\eta\r)\,+\, \mathrm{C}_{2}\,\mathrm{H}^{^{(2)}}_{\nu}\l(-c_{_s}k\eta\r)\r]~,$$
where $\mathrm{C}_{1}$ and $\mathrm{C}_{2}$ are constants of integration, $\mathrm{H}^{^{(1)}}_{\nu}\l(x\r)$ and $\mathrm{H}^{^{(2)}}_{\nu}\l(x\r)$ are Hankel functions of first and second kind, respectively. On imposing the Bunch-Davis initial condition that $u_{_k} = (2kc_{_s})^{-(1/2)}\exp\l[-ic_{_s}k\eta\r]$ at the sub-horizon scales $(\,-c_{_s}k\eta\,\gg\, 1\,)$ leads to  $\mathrm{C}_{2} = 0$ and the above solution for the mode function $u_{_k}$ becomes
\beq
u_{_k}\,=\, \sqrt{\frac{-\pi \eta}{4}}\,\exp\l[i\l(\frac{\pi}{2}\r)\l(\nu + \frac{1}{2}\r)\r]\,\mathrm{H}^{^{(1)}}_{\nu}\l(-c_{_s}k\eta\r)~.
\label{eqn: uk GPLI}
\eeq
Similarly for the tensor perturbations, the solution of the Eq.~(\ref{eqn: MS eqn tensor k-inf}), satisfying the Bunch-Davis initial condition turns out to be identical to $u_{_k}$, except for tensor perturbations $c_{_s} = 1$. Therefore,
\beq
v_{_k}\,=\, \sqrt{\frac{-\pi \eta}{\;4}}\,\exp\l[i\l(\frac{\pi}{2}\r)\l(\nu + \frac{1}{2}\r)\r]\,\mathrm{H}^{^{(1)}}_{\nu}\l(-k\eta\r)~.
\label{eqn: vk GPLI}
\eeq
From the above solutions for mode functions $u_{_k}$ and $v_{_k}$, the scalar and tensor power spectrum defined in Eqs.~(\ref{eqn: scalar PS}) and (\ref{eqn: tensor PS}), respectively, at the super horizon scales $(\,-c_{_s}k\eta\,\ll\, 1\,)$ turns out to be
\ber
\mathcal{P}_{_{S}}(k) &=& A_{_S}\,k^{-2/(q-1)}~,
\label{eqn: scalar PS GPLI}\\
\mathcal{P}_{_{T}}(k) &=& A_{_T}\,k^{-2/(q-1)}~,
\label{eqn: tensor PS GPLI}
\eer
where
\ber
A_{_S} &=& \l(\frac{q\,\lambda^{2}}{16\,\pi^{3}\,c_{_s}^{2\nu - 2}}\r)
\l(\frac{\Gamma(\nu)\,2^{\nu}}{\mpl\,a_{_1}}\r)^{2}~,
\label{eqn: As GPLI}\\
A_{_T} &=& \l(\frac{1}{\pi^{3}}\r)
\l(\frac{\Gamma(\nu)\,2^{\nu}}{\mpl\,a_{_1}}\r)^{2}~.
\label{eqn: At GPLI}
\eer
In the above equations $\Gamma(\nu)$ is Gamma function and $a_{_1}$ is a constant that appears in the equation describing the evolution of scalar factor in conformal time, \emph{viz}.\ $a(\eta) =  a_{_1}\l[-\eta\r]^{-q/(q - 1)}$.
The value of this parameter $a_{_1}$ in Eqs.~(\ref{eqn: As GPLI}) and (\ref{eqn: At GPLI}) can be fixed using the CMB normalization, namely, $P_{_{S}}(k_{\ast}) = 2.2\times10^{-9}$ at the pivot scale $k_{\ast} = 0.05\,\mathrm{Mpc}^{-1}$~\cite{Planck-inflation}. For $n = 4$ and $q = 60$, it turns out that $a_{_1} \simeq 4.4\times10^{5}\mpl^{-q/(q - 1)}$. Since $a(\eta) =  a_{_1}\l[-\eta\r]^{-q/(q - 1)}$ implies that $a(t) = a_{\ast}(t/t_{\ast})^{q}$, it turns out that  $a_{\ast} = a_1^{1-q}\l[t_\ast/(q-1)\r]^{q}$.  Let $t_{\ast} = q/H_{\ast}$ be the time at which the pivot scale exit the cosmological horizon ($a_{\ast}H_{\ast} = k_{\ast}$), which in turn implies that
\ber
H_{\ast} = \frac{1}{a_{_1}}\l[\l(\frac{1}{k_{\ast}}\r)\l(\frac{q}{q - 1}\r)^{q}\r]^{\frac{1}{q - 1}}~.\nn
\eer
Therefore, the CMB normalized value of $a_{_1}$ can be used to determined the value of the Hubble parameter $H_{\ast}$ when the pivot scale exit the cosmological horizon. For $n = 4$ and $q = 60$, we find that $H_{\ast} = 2.2\times10^{-5}\mpl$.

From Eqs.~(\ref{eqn: scalar PS GPLI}) and  (\ref{eqn: tensor PS GPLI}), the scalar and tensor spectral index, defined in Eqs.~(\ref{eqn: ns definition}) and (\ref{eqn: nt definition}), respectively, turns out to be
\beq
n_{_{S}} - 1\,=\,n_{_{T}}\,=\,-\frac{2}{q - 1}~.
\label{eqn: ns nt GPLI}
\eeq
This is exactly the same for the case of standard power law inflation driven by a canonical scalar field with an exponential potential~\cite{Abbott-1984,Lucchin-1985,sahni88} and one also gets the same $n_{_{S}}$ and $n_{_{T}}$ in the power law scenario in some non-canonical scalar field models such as those discussed in Refs.~\cite{Garriga-1999,Picon-1999,sanil-2013}.
In fact, Eq.~(\ref{eqn: ns nt GPLI}) is valid for any model of power law inflation  based on the Lagrangian~(\ref{eqn: G-Lagrangian}) but for which the parameters $c_{_s}^{2}$ and $\lambda$ defined in Eqs.~(\ref{eqn: sound speed G-inf def}) and (\ref{eqn: lambda def}), respectively, are identically constant.
This simply follows from the fact that whenever $c_{_s}^{2}$ and $\lambda$ are constant, the solution~(\ref{eqn: uk GPLI}) for the mode function $u_{_k}$ satisfy the Mukhanov Sasaki equation~(\ref{eqn: MS eqn G-inf}) during power law expansion and the resultant $n_{_{S}}$ (and $n_{_{T}}$) is the consequence of this solution for $u_{_k}$ (and $v_{_k}$).

Eqs.~(\ref{eqn: scalar PS GPLI}) and  (\ref{eqn: tensor PS GPLI}) lead to the following tensor-to-scalar ratio
\beq
r\,  =\, \l(\frac{16}{q\,\lambda^{2}}\r)\l(\frac{1}{c_{_s}}\r)^{\frac{1 + q}{1 - q}}~.
\label{eqn: T-to-S GPLI}
\eeq
It is evident from the above expression that unlike the spectral indices $n_{_{S}}$ and $n_{_{T}}$ which depends only on the value of $q$ in the power law solution $a(t) \propto t^q$, the tensor-to-scalar ratio $r$ also depends on the model parameters $c_{_s}^{2}$ and $\lambda$ defined in Eqs.~(\ref{eqn: sound speed G-inf def}) and (\ref{eqn: lambda def}), respectively.  Therefore,  $r$ contains the details of the dynamics of inflation and it can play an important role in distinguishing models of inflation~\cite{A-Majumdar-T-to-S}.

Eqs.~(\ref{eqn: ns nt GPLI})  and (\ref{eqn: T-to-S GPLI}) imply the following consistency relation
\beq
r\,  =\, -\l(\frac{8\,n_{_{T}}}{\lambda^{2}}\r)\l(\frac{q - 1}{q}\r)\l(\frac{1}{c_{_s}}\r)^{\frac{1 + q}{1 - q}}~,
\label{eqn: consistency reln GPLI}
\eeq
where the expression for $c_{_s}^{2}$ and $\lambda$  for the Galilean power law inflation model~(\ref{eqn: Lagrangian GPLI}) are given by Eqs.~(\ref{eqn: sound speed GPLI}) and (\ref{eqn: lambda GPLI}), respectively.
Note that the above consistency relation is an exact result since no slow roll approximation is imposed.

\begin{figure}[t]
\begin{center}
\scalebox{0.83}[1]{\includegraphics{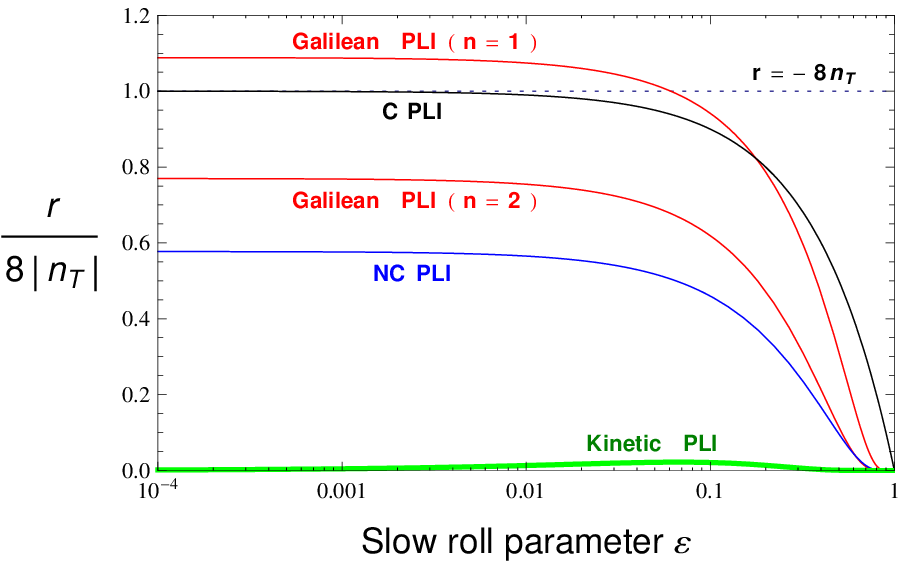}}
\scalebox{0.83}[1]{\includegraphics{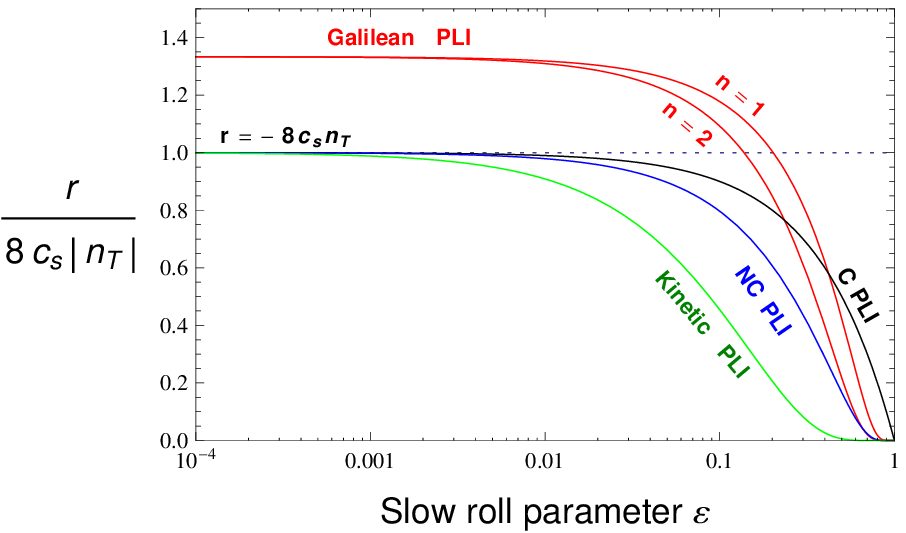}}
\caption{In the left panel, the ratio $r/(8|n_{_{T}}|)$ is plotted as function of the slow roll parameter $\varepsilon = -\dot{H}/(H^{2})$ for the following four power law inflation models: $(i)$ Galilean PLI: Power law inflation in Galilean model~(\ref{eqn: Lagrangian GPLI}) with $n = 1$ and $n = 2$, $(ii)$ C PLI: Canonical scalar field driven power law inflation, $(iii)$ NC PLI: A non-canonical scalar field model of power law inflation describe by the Lagrangian~(\ref{eqn: Lagrangian NPLI}) with $\alpha = 2$ and $(iv)$ Kinetic PLI: A kinetic driven power law inflation model~\cite{Garriga-1999,Picon-1999} where ${\cal L} = f(\phi)\l(-X\,+\,X^{2}\r)$ and $f(\phi) \propto \phi^{-2}$. In the right panel, the ratio $r/(8c_{_s}|n_{_{T}}|)$ is plotted for the same set of power law inflation models. Note that, in the slow roll limit, the ratio $r/(8c_{_s}|n_{_{T}}|)$ approaches unity for canonical, kinetic and non-canonical models whereas, for the Galilean  model of power law inflation, this ratio approaches $4/3$ irrespective of the value of $n$ in the Lagrangian~(\ref{eqn: Lagrangian GPLI}).}
\label{fig: consistency reln}
\end{center}
\end{figure}

In Fig.~\ref{fig: consistency reln}, the ratios $r/(8|n_{_{T}}|)$ and  $r/(8c_{_s}|n_{_{T}}|)$ are plotted as a function of slow roll parameter $\varepsilon \equiv -\dot{H}/H^{2}$ for different models of power law inflation. In this figure, relation~(\ref{eqn: consistency reln GPLI}) is used for the Galilean power law model while  for the power law scenario in k-inflation~\cite{Picon-1999} and  in the non-canonical scalar field model~\cite{sanil-2013}, the same relation~(\ref{eqn: consistency reln GPLI}) is used but with $\lambda = 1$. This is justified from the analysis of Ref.~\cite{Garriga-1999}.
For the power law inflation driven by a canonical scalar field  $r = 16/q$~\cite{Tarun-92} which leads to the following exact consistency relation $r = -8n_{_{T}}(q - 1)/q$.

In the slow roll limit which corresponds to $q \gg 1$, the consistency relation~(\ref{eqn: consistency reln GPLI}) reduces to
\beq
r\,  \simeq\, -\frac{8\,c_{_s}\,n_{_T}}{\lambda^{2}}~.
\label{eqn: consistency reln G-inf SR limlt}
\eeq
In comparison, for the case of k-inflation, one gets $r\,  =\,- 8\,c_{_s}\,n_{_T}$. Therefore, as mentioned earlier, it is the $\lambda$ parameter defined in Eq.~(\ref{eqn: lambda def}) which alters the consistency relation in G-inflation.
Note that, although, the consistency relation~(\ref{eqn: consistency reln G-inf SR limlt}) was derived for the power law inflation, it is approximately valid, in the slow roll limit,  for a generic G-inflation model with Lagrangian~(\ref{eqn: G-Lagrangian}).
For the power law inflation driven by a Galilean field, the slow roll regime of the consistency relation~(\ref{eqn: consistency reln G-inf SR limlt}) can be re-expressed as
\beq
r\,  \simeq\, -8\,n_{_T}\l(\frac{4}{9}\r)\sqrt{\frac{6}{n}}~.
\label{eqn: consistency reln GPLI SR limlt}
\eeq
When $n = 1$, the above relation becomes $r \simeq -(32/9)\sqrt{6}\;n_{_T}$ consistent with those derived in Ref.~\cite{Kamada-2010}.
For any value of $n$, Eq.~(\ref{eqn: consistency reln GPLI SR limlt}) can also be expressed as $r \simeq (64/9)\,\l(\sqrt{6/n}\,\r) \varepsilon$, where $\varepsilon$ is the slow roll parameter, which for the power law solution $a(t) \propto t^{q}$  turns out to be $\varepsilon\,=\, q^{-1}$.
Although, we considered the Galilean power law inflation model~(\ref{eqn: Lagrangian GPLI}) for which $G(X,\phi) \propto X^{n}$, the expression  $r \simeq (64/9)\,\l(\sqrt{6/n}\,\r)\varepsilon$  is also valid in case of Higgs G-inflation model where $G(X,\phi)  \propto \phi^{2n + 1}X^{m}$ as described in Ref.~\cite{Kamada-2013}.
However, recall that the expression  $r \simeq (64/9)\,\l(\sqrt{6/n}\,\r)$ is slow roll limit of the exact relation~(\ref{eqn: consistency reln GPLI}).

When $n = 1$, Eq.~(\ref{eqn: consistency reln GPLI SR limlt}) implies that $r > - 8 n_{_T}$ even though the speed of sound is subluminal $(c_{_s}^{2} = 2/3)$.
This does not arise in the case of canonical and non-canonical scalar field models of inflation for which $r \leq - 8 n_{_T}$ when $c_{_s}^{2} \leq 1$~\cite{Garriga-1999}. This leads us to ask the following important question: What is the physical reason behind the violation of the standard consistency relation in G-inflation models~?
To go about answering this question let us re-write the scalar power spectrum~(\ref{eqn: scalar PS GPLI}) as
\beq
\mathcal{P}_{_{S}}(k) = \lambda^{2}\; ^{^{(nc)}}\mathcal{P}_{_{S}}(k)~,
\label{eqn: scalar PS GPLI vs KPLI}
\eeq
where $^{^{(nc)}}\mathcal{P}_{_{S}}(k)$  is the scalar power spectrum that one gets in an equivalent non-canonical model of inflation which leads to the same background evolution, \emph{viz}.\  $a(t) \propto t^q$ and for which the value of $c_{_s}^{2}$ is the same as given in Eq.(\ref{eqn: sound speed GPLI}).
Note that the expression for the tensor power spectrum~(\ref{eqn: tensor PS GPLI}) remains unchanged for an equivalent non-canonical scalar model, as tensor perturbations do not directly couple with scalar perturbations.
In the slow roll limit the expression~(\ref{eqn: lambda GPLI}) implies that $\lambda \simeq \sqrt{3/4}$. therefore in the slow roll limit Eq.~(\ref{eqn: scalar PS GPLI vs KPLI}) implies that $\mathcal{P}_{_{S}}(k)\,<\, ^{^{(nc)}}\mathcal{P}_{_{S}}(k)$. This means that the scalar power spectrum in G-inflation models is suppressed by a factor $\lambda^{2}$ as compared to the same in an equivalent k-inflation scenario and consequently this enhances the tensor-to-scalar ratio. It is because of this enhancement of  tensor-to-scalar ratio, one can get $r > - 8 n_{_T}$  in G-inflation models even when $c_{_s}^{2} \leq 1$.

Note that for integer value of $n$, it follows from Eq.~(\ref{eqn: consistency reln GPLI SR limlt}) that $r > - 8 n_{_T}$ only when $n = 1$. For $n \geq 2$ one gets back the standard consistency relation $r < - 8 n_{_T}$ as in the case of inflation driven by a canonical or non-canonical scalar field.

From Eqs.~(\ref{eqn: sound speed GPLI}), (\ref{eqn: lambda GPLI}) and (\ref{eqn: consistency reln GPLI}), it follows that, if we restrict $c_{_s}^{2} \leq 1$, then
\beq
\frac{r}{8\,|n_{_T}|} \,\leq\, \frac{4}{3}~.
\label{eqn: r by nt limit}
\eeq
Therefore, the upper bound on the ratio $r/(8\,|n_{_T}|)$ for the Galilean power law model~(\ref{eqn: Lagrangian GPLI}) is $1.333$, when the speed of sound for the scalar field perturbations is restricted to be subluminal.
This is the main result of this paper. In comparison, for all canonical and non-canonical scalar field models of inflation, the upper bound on the ratio $r/(8\,|n_{_T}|)$ is unity when $c_{_s}^{2} \leq 1$.

Although the expression~(\ref{eqn: r by nt limit}) was obtained for a restricted class of G-inflation model~(\ref{eqn: Lagrangian GPLI}) driving power law inflation, it may be worthwhile to investigate whether the inequality $r/(8\,|n_{_T}|)\, \leq \,4/3$ holds for a generic G-inflation model with the Lagrangian~(\ref{eqn: G-Lagrangian}) but for which $c_{_s}^{2} \leq 1$.

\subsection{Observational viability of Galilean power law inflation}

\begin{figure}[t]
\begin{center}
\scalebox{1.1}[1.2]{\includegraphics{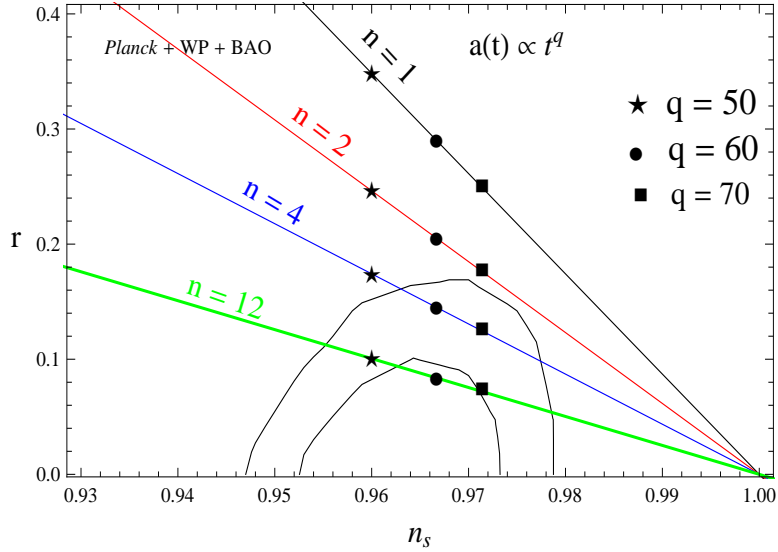}}
\caption{The constraints from the CMB observations on the power law inflation driven by a Galilean field with the Lagrangian~(\ref{eqn: Lagrangian GPLI}) and with an inverse power law potential~(\ref{eqn: V-phi GPLI}). For different values of the parameter $n$ in (\ref{eqn: Lagrangian GPLI}),  the scalar spectral index $n_{_S}$ and the tensor-to-scalar ratio $r$ are shown for $q = 50$, $60$ and $70$ in the power law solution $a(t) \propto t^{q}$.  The inner and outer contours represents the  $1\sigma$ and $2\sigma$ confidence limits obtained using the CMB data from {\em Planck} along with the large angle polarization data from WMAP  in combination with the  Baryon acoustic oscillations data ({\em Planck}~$+$~WP~$+$~BAO)~\cite{Planck-inflation}.}
\label{fig:ns-r}
\end{center}
\end{figure}

Let us now confront the Galilean PLI model with the recent constraints from the CMB observation from  \emph{Planck} mission which indicate $n_{_{S}}\,=\,0.9603\,\pm\,0.0073$ and $r\,<\,0.11$ at $95\%$ CL~\cite{Planck-inflation}.
In fact, at $95\%$ CL, \emph{Planck} data allows $n_{_{S}}$ in the range $[0.945-0.98]$. For $n_{_{S}}$ to lie in that range, it follows from Eq.~(\ref{eqn: ns nt GPLI}) that $q$ in the power law  solution  $a(t) \propto t^q$ must be within the range $\sim [38 - 101]$.
Consequently, one finds from Eq.~(\ref{eqn: consistency reln GPLI SR limlt}) that $0.17\,\lesssim\,r\,\lesssim\,  0.46$  for $n = 1$, well above the bound set by the \emph{Planck} results which  indicate $r\,<\,0.11$ at $95\%$ CL~\cite{Planck-inflation}.
For example, when $n = 1$ and $q = 50$,  Eq.~(\ref{eqn: ns nt GPLI}) gives $n_{_{S}} \simeq 0.96$ consistent with  \emph{Planck}, however, Eq.~(\ref{eqn: consistency reln GPLI SR limlt}) gives a larger value of $r \simeq  0.35$. Recall that, it is for $n = 1$ the Galilean power law inflation model~(\ref{eqn: Lagrangian GPLI}) violates the standard consistency relation $r \leq - 8 n_{_T}$.

For $n = 2$ and $q = 100$, Eqs.~(\ref{eqn: ns nt GPLI}) and (\ref{eqn: consistency reln GPLI SR limlt}) gives $n_{_{S}} \simeq 0.98$ and $r \simeq 0.12$ marginally outside the \emph{Planck} bounds. Since $r \propto \l(n\r)^{-1/2}$ , larger values of $n$ lowers $r$ for a given value of $q$. For example, when $n = 8$ and $q = 60$, one gets $n_{_{S}} \simeq 0.966$ and $r \simeq 0.1$ well within the \emph{Planck} results which indicate $r\,<\,0.11$ at $95\%$ CL~\cite{Planck-inflation}.
Therefore, for $n > 2$, the Galilean power law inflation model~(\ref{eqn: Lagrangian GPLI}) gives observationally consistent value for $n_{_{S}}$ and $r$ (see Fig.~\ref{fig:ns-r}). However, recall that for $n > 2$, the model leads to $r < - 8 n_{_T}$.

Hence, although the Galilean power law inflation  model can lead to $r > - 8 n_{_T}$, in the domain of the model parameter space which gives observationally viable results, the model respects the standard consistency relation $r \leq - 8 n_{_T}$.

\section{Comparing Galilean and non-canonical scalar inflation}
\label{sec: GPLI vs NCPLI}
\subsection{In the case of power inflation}
Since for an observationally viable Galilean power law inflation model $r < - 8 n_{_T}$, it may be possible to map this model to an equivalent  scenario in a non-canonical scalar field settings\footnote{It is important to note that G-inflation models~(\ref{eqn: G-Lagrangian}) with $G_{X} \neq 0$ and  k-inflation models are fundamentally different scenarios. By the statement: ``G-inflation power law model can be mapped to an equivalent scenario in a k-inflation model", all that we mean is that the two models leads to same set of CMBR observables $n_{_S}$, $n_{_T}$ and $r$.} which necessarily gives $r < -8n_{_T}$  when $c_{_s}^{2} < 1$.
It is not possible to do so with the standard power law inflation driven by canonical scalar field with an exponential potential since for canonical inflation $r = - 8 n_{_T}$ and moreover the value of tensor-to-scalar ratio in such a model turns out to be larger than the bound set by the \emph{Planck} results~\cite{Planck-inflation}.

In the Galilean power law  model~(\ref{eqn: Lagrangian GPLI}), the speed of sound as given in Eq.~(\ref{eqn: sound speed GPLI}) approaches a constant value in the slow roll limit and it is independent of the value of $q$ in the power law solution $a(t) \propto t^q$.
In contrast, in the kinetic driven power law inflation model~\cite{Garriga-1999, Picon-1999} with ${\cal L} = f(\phi)\l(-X\,+\,X^{2}\r)$, where $f(\phi) \propto \phi^{-2}$, the speed of sound $c_{_s}$ depends on the value of $q$ and in the slow roll limit one finds that $c_{_s}^{2} \simeq 1/(12 q)$.
Mapping the Galilean power law  model~(\ref{eqn: Lagrangian GPLI}) onto the kinetic driven model depends on the value of $q$ in the power law  solution  $a(t) \propto t^q$. We therefore consider the following non-canonical model of inflation~\cite{sanil-2012,Li-2012}
\beq
{\cal L}(X,\phi) = X\l(\frac{X}{M^{4}}\r)^{\alpha-1} -\; V(\phi),
\label{eqn: Lagrangian NPLI}
\eeq
where the parameter $\alpha$ is  dimensionless  while $M$ has  mass dimension of one.
The above Lagrangian can be viewed as a generalization of the standard canonical scalar field Lagrangian  which corresponds to setting $\alpha = 1$ in Eq.~(\ref{eqn: Lagrangian NPLI}). Note that the structure of the restricted class of Galilean inflation model~(\ref{eqn: Lagrangian GPLI}) is similar the above Lagrangian and astonishingly in this model also an inverse power law potential of the form~\cite{sanil-2013}
\beq
V(\phi)  = \frac{V_{_0}}{(\phi/\mpl)^{s}}~;~~~~\mathrm{where}~~~s  = \frac{2\,\alpha}{\alpha - 1}~,
\label{eqn: V-phi}
\eeq
can drive power law inflation. The speed of sound in the model~(\ref{eqn: Lagrangian NPLI}) turns out to be constant and is given by
\beq
c_{_S} = \f{1}{\sqrt{2\,\alpha - 1}}.
\label{eqn: sound speed NPLI}
\eeq
Since $c_{_S}$ is also constant and independent of $q$ in the slow roll limit in Galilean power law inflation model, mapping between the two models is possible.

In the non-canonical power law inflation model~(\ref{eqn: Lagrangian NPLI}), the spectral indices $n_{_S}$  and $n_{_T}$ are exactly the same as those in the Galilean model given in Eq.~(\ref{eqn: ns nt GPLI}). However, the tensor-to-scalar ratio is different and is given by
\beq
r\, \simeq\, \f{16}{q\,\sqrt{2\,\alpha - 1}}~.
\label{eqn: T-to-S NPLI}
\eeq
The power law inflation  model~(\ref{eqn: Lagrangian NPLI}) gives observationally viable values for $n_{_S}$ and $r$ for $\alpha \geq 2$ as per the recent \emph{Planck} results~\cite{sanil-2013}.
For a given background evolution $a(t) \propto t^q$, the two power law inflation models~(\ref{eqn: Lagrangian GPLI}) and (\ref{eqn: Lagrangian NPLI}) gives the same set of value for the observables $\l\{n_{_S},\, n_{_T}, \,r \r\}$, in the slow roll limit, if
\beq
n\,=\, \l(\frac{32}{27}\r)\l(2\alpha - 1\r).
\label{eqn: alpha n}
\eeq
In other words, the two power law inflation  models are observationally indistinguishable \emph{in the slow roll limit} as far as the observational parameters $\l\{n_{_S},~ n_{_T}, ~r \r\}$  are concerned, if the above relation is satisfied.
However, it is possible that the non-gaussianity parameter $f_{_{\mathbf{NL}}}$~\cite{Planck-NG} can act as a discriminator between the two models~\cite{Kobayashi-2011b}.

Although the two power law inflation models~(\ref{eqn: Lagrangian GPLI}) and (\ref{eqn: Lagrangian NPLI}) give the same values for $\l\{n_{_S},\, n_{_T}, \, r \r\}$, it is important to note that the speed of sound is different in both the models.
In fact, for the equivalent non-canonical power law inflation model~(\ref{eqn: Lagrangian NPLI}), the speed of sound turns out to be $c_{_s}\times\lambda^{-2}~,$ where $c_{_s}$ and $\lambda$  for the Galilean power law inflation model~(\ref{eqn: Lagrangian GPLI}) are given in Eqs.~(\ref{eqn: sound speed GPLI}) and (\ref{eqn: lambda GPLI}), respectively.
It is important  to note that this is generically true in the slow roll limit for  any Galilean inflation model based on the Lagrangian~(\ref{eqn: G-Lagrangian}), which means that a Galilean model~(\ref{eqn: G-Lagrangian}) characterized by the parameters $c_{_s}$ and $\lambda$, defined in Eqs.~(\ref{eqn: sound speed G-inf def}) and (\ref{eqn: lambda def}), respectively, can be mapped to an equivalent k-inflation model whose speed of sound is $c_{_s}\times\lambda^{-2}$.
In such a scenario, the non-canonical scalar field model gives the same value for the observables $\l\{n_{_S},\, n_{_T}, \, r \r\}$ as in the case of Galilean inflation.
The proof the above statement is given below.

\subsection{Generic slow roll G-inflation model}

Let us consider a generic Galilean model of inflation described by the Lagrangian~(\ref{eqn: G-Lagrangian}).
From the Mukhanov-Sasaki equation~(\ref{eqn: MS eqn G-inf}), it is clear that at scales much larger than the sound horizon (\ie.  at $a H >> c_{_s} k$), the mode function $u_{_k} \simeq A_{_k} z$, where $A_{_k}$ is the $k$ dependent constant of integration.
However, at scales much below the sound horizon (\ie.  at $a H << c_{_s} k$), the Bunch-Davis initial condition implies that $u_{_k} = (2 c_{_s} k)^{-1/2}\exp [ - i k c_{_s} \eta]$.
The constant $A_{_k}$ can, therefore, be fixed by equating the two regimes of solution for $u_{k}$ at the sound horizon  $a H = c_{_s} k$. Consequently one finds that at scales much above the sound horizon
\beq
\l|\frac{u_{_k}}{z}\r|^{2}\,=\, \l(\frac{1}{2 k c_{_s} z^{2}}\r)_{_{a H = c_{_s} k}}~,
\eeq
where the quantity on the right hand side must be evaluated when the mode leaves the sound horizon.
Substituting the above solution for $u_{_k}$ in Eq.~(\ref{eqn: scalar PS}) gives
\beq
\mathcal{P}_{_{S}}(k)\,=\, \l(\frac{\lambda^{2}\,H^{4}}{4\, \pi^{2}\, c_{_s} \l(\rho_{_\phi} + p_{_\phi}\r)\,}\r)_{_{a H = c_{_s} k}}~,
\label{eqn: scalar ps g-inf SR limit}
\eeq
where $c_{_s}$ and $\lambda$ are defined in Eqs.~(\ref{eqn: sound speed G-inf def}) and (\ref{eqn: lambda def}), respectively.

Similarly for tensor perturbations
\beq
\mathcal{P}_{_{T}}(k)\,=\, \l(\frac{2\,H^{2}}{\pi^{2}\, \mpl^{2}}\r)_{_{a H =  k}}~.
\label{eqn: tensor ps g-inf SR limit}
\eeq
Note that unlike the case of scalar power spectrum~(\ref{eqn: scalar ps g-inf SR limit}) which is evaluated at the sound horizon $a H = c_{_s} k$, the above expression for tensor power spectrum must be evaluated when the modes leaves the horizon $a H  = k$.

During slow roll inflation, $H\,\simeq\,$ constant. In addition if the variation of $c_{_s}$ in the time scale of an e-fold of expansion is small enough, it is reasonable to approximate
\beq
\d\, \mathrm{ln}\, k\,\simeq \,\d\,\mathrm{ln}\, a~,
\label{eqn: d lnk dln a}
\eeq
both at the sound horizon $a H = c_{_s} k$ and at the horizon $a H  = k$.
Consequently one finds from Eqs.~(\ref{eqn: scalar ps g-inf SR limit}) and (\ref{eqn: tensor ps g-inf SR limit}) that \ber
n_{_{S}} - 1\, &\simeq&\, -4\varepsilon + 2\delta - \sigma_{_s} + 2\sigma_{_\lambda} ~,\label{eqn: ns g-inflation}\\
n_{_{T}}\, &\simeq&\, -2\varepsilon~,\label{eqn: nt g-inflation}
\eer
where $\varepsilon$, $\delta$, $\sigma_{_s}$ and $\sigma_{_\lambda}$ are the first order slow roll parameters defined as
\ber
\varepsilon &\equiv& -\frac{\dot{H}}{\,H^{2}}~,\label{eqn: SR pram 1} \\
\delta &\equiv& \varepsilon -  \frac{\dot{\varepsilon}}{2\,H\,\varepsilon}~,\label{eqn: SR pram 2}\\
\sigma_{_s} &\equiv& \frac{\dot{c_{_s}}}{H\,c_{_s}}~,\label{eqn: SR pram 3}\\
\sigma_{_\lambda} &\equiv& \frac{\dot{\lambda}}{H\,\lambda}~.\label{eqn: SR pram 4}
\eer
These slow roll parameters are generic to any scalar field models of inflation based on the Lagrangian~(\ref{eqn: G-Lagrangian}). Depending on the form of the functions $K(X,\phi)$ and $G(X,\phi)$, it is possible to define model specific slow roll parameters, see for instance~\cite{G-inf 1st paper,Kamada-2010}.

Although the two power spectrums~(\ref{eqn: scalar ps g-inf SR limit}) and (\ref{eqn: tensor ps g-inf SR limit}) are evaluated at two different moments in time, with the scalar power spectrum evaluated when the mode crosses the sound horizon while  the tensor power spectrum evaluated when the mode leaves the horizon,  one approximately gets the same results in the slow roll limit, even if both are evaluated at the horizon crossing $a H =  k$.
Therefore, it follows from Eqs.~(\ref{eqn: scalar ps g-inf SR limit}) and (\ref{eqn: tensor ps g-inf SR limit}) that
\beq
r\, =\, \frac{16\,c_{_s}\,\varepsilon}{\lambda^{2}}~.
\label{eqn: r g-inf SR limit}
\eeq
Note that the expression for $n_{_{S}}$, $n_{_{T}}$ and $r$ as described in Eqs.~(\ref{eqn: ns g-inflation}), (\ref{eqn: nt g-inflation}) and (\ref{eqn: r g-inf SR limit}), respectively, are valid in the slow roll limit of G-inflation model~(\ref{eqn: G-Lagrangian}). In the case of further generalization of G-inflation scenario which contains K-inflation models as its subclass, a detailed description of the primordial power spectra and their consequent observables $n_{_{S}}$, $n_{_{T}}$ and $r$ can be found in Ref.~\cite{Kobayashi-2011a}.

Let us now compare these values for $n_{_{S}}$, $n_{_{T}}$ and $r$ for non-canonical scalar field models of inflation. For consistency in notation, let the Lagrangian of the non-canonical scalar field be represented as
\beq
{\cal L}\,=\,\widetilde{K}(X,\phi)
\label{eqn: K Lagrangian}
\eeq
In this model, the squared speed of sound turns out to be\footnote{The notation  $\widetilde{c_{_s}}$ is used here to describe the speed of sound in k-inflation just to differentiate it with $c_{_s}$ which represents the speed of sound in G-inflation.}
\beq
\widetilde{c_{_s}}^{2} \equiv  \f{\widetilde{K}_{X}}{\widetilde{K}_{X}\, + \, 2X \widetilde{K}_{XX}}~.
\label{eqn: sound speed k-inf def}
\eeq
In a generic k-inflation model~(\ref{eqn: K Lagrangian}), it turns out that in the slow roll limit~\cite{Garriga-1999}
\ber
n_{_{S}} - 1\, &\simeq&\, -4\varepsilon + 2\delta - \widetilde{\sigma}_{_s}~,\label{eqn: ns k-inflation}\\
n_{_{T}}\, &\simeq&\, -2\varepsilon~,\label{eqn: nt k-inflation}\\
r\, &\simeq&\, 16\,\widetilde{c}_{_s}\,\varepsilon~,\label{eqn: r k-inflation}
\eer
where $\varepsilon$ and $\delta$ are defined in Eqs.~(\ref{eqn: SR pram 1}) and (\ref{eqn: SR pram 2}), respectively, and $\widetilde{\sigma}_{_s} \equiv \dot{\widetilde{c}_{_s}}/(\widetilde{c}_{_s} H)$ is the same as those defined in Eq.~(\ref{eqn: SR pram 3}), the only difference here is that $c_{_s}$ in Eq.~(\ref{eqn: SR pram 3}) is replaced by $\widetilde{c}_{_s}$.

Let the k-inflation model~(\ref{eqn: K Lagrangian}) be such that it lead to the same background evolution for $a(t)$ as those in the G-inflation model~(\ref{eqn: G-Lagrangian}). In addition, if its speed of sound is such that
\beq
\widetilde{c_{_s}}\, =\, \frac{c_{_s}}{\lambda^{2}}~,\label{eqn: cs tilde cs reln}
\eeq
where $c_{_s}$ and $\lambda$ are defined in Eqs.~(\ref{eqn: sound speed G-inf def}) and (\ref{eqn: lambda def}), respectively, it follows from  equations (\ref{eqn: ns g-inflation}), (\ref{eqn: nt g-inflation}), (\ref{eqn: r g-inf SR limit}) and  (\ref{eqn: ns k-inflation}) to (\ref{eqn: r k-inflation}) that the observables $\l\{\,n_{_{S}},\, n_{_{S}},\, r\,\r\}$ are the same for both the models~(\ref{eqn: G-Lagrangian}) and (\ref{eqn: K Lagrangian}).

Hence, in the slow roll regime of any generic G-inflation model~(\ref{eqn: G-Lagrangian}) one can associate a k-inflation model~(\ref{eqn: K Lagrangian}) whose speed of sound satisfy the relation~(\ref{eqn: cs tilde cs reln}). The two models will then be observationally indistinguishable as far as the basic set of observables $\l\{\,n_{_{S}},\, n_{_{S}},\, r\,\r\}$ are concerned.

In the case of k-inflation models, the non-gaussianity parameter in the equilateral limit is inversely proportional to the square of the speed of sound $c_{_s}^{2}$  whereas the tensor-to-scalar ratio $r$ is proportional to $c_{_s}$~\cite{chen-2007}. Therefore, for a given value of the slow roll parameter $\varepsilon$, k-inflation models with smaller $r$ leads to larger $f_{_{\mathbf{NL}}}^{\mathrm{equil}}$. However, in the case of G-inflation models, it is possible to have a large value of  $f_{_{\mathbf{NL}}}^{\mathrm{equil}}$ even when $r$ is large~\cite{Kobayashi-2011b}. Therefore, the non-gaussianity parameter $f_{_{\mathbf{NL}}}$ along with the other observables such as $r$ can, in principle, discriminate between the G-inflation and k-inflation scenarios~\cite{Gao-2013,Kamada-2013b}.
Note that the recent CMB observation from \emph{Planck} has indicated that $f_{_{\mathbf{NL}}}^{\mathrm{equil}}\, =\, -42\,\pm\,75$~\cite{Planck-NG} whereas $r\,<\,0.11$ at $95\%$ CL~\cite{Planck-inflation}.

It is also important to note that in Galilean models of inflation an additional slow roll parameter $\sigma_{_\lambda}$, defined in Eq.~(\ref{eqn: SR pram 4}), is introduced to describe the slow roll inflation.
The slow roll approximation, at the leading order, not only corresponds to the conditions $\varepsilon << 1$ and $|\delta| << 1$, but also requires the following assumptions $|\sigma_{_s}| << 1$ and $|\sigma_{_\lambda}| << 1$.
In the scenario when the slow roll parameters and their time variations are not small enough, it may be worth pursuing whether the generalised slow roll approach for calculating the power spectrums can be extended to Galilean models of inflation~\cite{Stewart-2002,Hu-2011}.

\section{Summary and conclusions}
\label{sec: conclusions}
In this paper we considered a restricted class of Galilean inflation models with the Lagrangian of the form~(\ref{eqn: Lagrangian GPLI}) and showed that an inverse power law potential~(\ref{eqn: V-phi GPLI}) can lead to power law inflation. An exact inflationary consistency relation was derived for this model without imposing the slow roll condition and its evolution with the slow roll parameter $\varepsilon$ is depicted in Fig.~\ref{fig: consistency reln}.
From the consistency relation, it turns out that one can have both $r > - 8 n_{_T}$ or $r \leq - 8 n_{_T}$, depending on the value of the parameter $n$ in the Lagrangian~(\ref{eqn: Lagrangian GPLI}), in spite of the fact that the speed of sound is subluminal in both the cases.
Interestingly, as the value of slow roll parameter $\varepsilon$ is varied from $\varepsilon << 1$ to $\varepsilon \simeq 1$, the consistency relation in the model which leads to  $r > - 8 n_{_T}$  evolves towards $r \leq - 8 n_{_T}$. This important result is depicted in Fig.~\ref{fig: consistency reln}. Therefore, the violation of the standard consistency relation ($r \leq - 8 n_{_T}$) in power law G-inflation model happens only in the slow roll regime ($\varepsilon << 1$). Hence for investigating the violation of consistency relation in a generic G-inflation model, it is highly reasonable to impose the slow roll approximation as done in Refs.~\cite{Kamada-2010,Kamada-2013,Kobayashi-2011a}.

In the slow roll regime of power law G-inflation model~(\ref{eqn: Lagrangian GPLI}), when the parameter $n \geq 2$  it turns out that $r < - 8 n_{_T}$, but for $n = 1$ the model leads to $r > - 8 n_{_T}$ even though $c_{_s} < 1$.
This is contrary to what one gets in inflation models based on a canonical or non-canonical scalar field for which $c_{_s} \leq 1$ necessarily implies that
$r \leq - 8 n_{_T}$.
We have identified the reason for this distinct behavior in Galilean inflation models and is  attributed to the fact that in these models the scalar power spectrum is suppressed by a factor $\lambda^{2}$, where $\lambda$ is defined in Eq.~(\ref{eqn: lambda def}), as compared to a k-inflation model which leads to the same background evolution and has the same speed of sound for scalar field perturbations.
Consequently, the tensor-to-scalar ratio is enhanced  by a factor  $\lambda^{-2}$ which alters the inflationary consistency relation which one gets in an equivalent k-inflation model.
On restricting the speed of sound in the Galilean power law inflation model~(\ref{eqn: Lagrangian GPLI}) to be subluminal, we find that the upper bound on the ratio $r/(8 |n_{_T}|)$ is $1.333$. This is the core result of this paper.

Even though it is possible to get $r > - 8 n_{_T}$ in the Galilean power law inflation model, the domain of the parameter $n$ which gives observationally viable value for $n_{_S}$ and $r$, as per the recent \emph{Planck} results, naturally lead to $r < - 8 n_{_T}$ which one gets in non-canonical scalar field model of inflation.
Therefore, as demonstrated in Sec.~\ref{sec: GPLI vs NCPLI}, one can map observationally viable Galilean model~(\ref{eqn: G-Lagrangian}) to  an equivalent non-canonical model which leads to the same background evolution but whose speed of sound satisfies relation~(\ref{eqn: cs tilde cs reln}). In such a scenario, the Galilean and the non-canonical scalar field model of inflation are observationally indistinguishable as per the basic  set of observables $\l\{n_{_S},\, n_{_T}, \, ~r \r\}$.

It is important to note that any generic k-inflation model is completely described by the behaviour of the equation of state parameter $w$ and the speed of sound $c_{_s}$. However, in the case of Galilean scalar field models~(\ref{eqn: G-Lagrangian}), an additional parameter $\lambda$ defined in Eq.~(\ref{eqn: lambda def}) is also required to describe the system completely. The reason for the appearance of this additional parameter $\lambda$ is related to the fact that Galilean scalar field models~(\ref{eqn: G-Lagrangian}) behaves as an imperfect fluid~\cite{Pujolas-2011} and it is well known that additional parameters besides $w$ and $c_{_s}$ are required to describe imperfect fluids.
For all standard single scalar field models, canonical or non-canonical, it turns out that $\lambda = 1$. In the case of inflation driven by a canonical scalar field, in addition to $\lambda = 1$, the speed of sound $c_{_s} = 1$. The non-gaussianity parameter $f_{_{\mathbf{NL}}}^{\mathrm{equil}}$, evaluated in the equilateral limit, can act as an observational signature of any deviation of the speed of sound from $c_{_s} = 1$~\cite{Planck-NG}. Similarly, it is  worth exploring whether there exists any observational consequences of deviation of the parameter $\lambda$ from $\lambda = 1$, which can then act as a definitive test of Galilean models of inflation.


\acknowledgments
We thank Misao Sasaki and Varun Sahni for enlightening discussions on inflationary consistency relation and also thank the support of Max Planck-India Partner Group on Gravity and Cosmology.
SS is partially supported by Ramanujan Fellowship of DST, India.

\appendix

\section{Speed of sound of scalar field perturbations}\label{sec: speed of sound}
Here, we derive the expression for the speed of sound with which the scalar field perturbations propagates in the Galilean model~(\ref{eqn: G-Lagrangian}). The equation of motion for the Galilean field in the curved space time is given in Eq.~(\ref{eqn: G_I EOM}). However, note that this equation contains the term $R_{\mu\nu}\partial^{\mu}\phi\partial^{\nu}\phi$. This term can be expressed solely in terms of the field $\phi$ and its derivatives since Einstein's equation implies that
\beq
R_{\mu\nu}\,=\,\l(\frac{1}{\mpl^{2}}\r)\l(T_{\mu\nu} - \frac{T}{2}g_{\mu\nu}\r)~.
\eeq
Substituting for $T_{\mu\nu}$ from Eq.~(\ref{eqn: EM tensor G-field}) in the above equation gives
\beq
R_{\mu\nu}\,\partial^{\mu}\phi\,\partial^{\nu}\phi\,=\,\l(\frac{2}{\mpl^{2}}\r)
\l[~X(K + XK_{X}) - 4X^{2}G_{\phi} + X^{2}G_{X}\Box \phi -2XG_{X}\partial^{\mu}\phi\,\partial_{\mu}X~\r]~.
\label{Aeqn: R mu nu}
\eeq
Substituting (\ref{Aeqn: R mu nu}) in Eq.~(\ref{eqn: G_I EOM}) we obtain the following equation of motion
\beq
B_{1}\,\Box\phi\, +\, B_{2}\,\partial^{\mu}\phi\,\partial_{\mu}X \,+\, B_{3}\,(2X)\,
 +\, B_{4}\,\,G_{X}\, + \, B_{5}\,G_{XX}\,+\,K_{\phi}\,=\,0~,
\label{Aeqn: G_I EOM}
\eeq
where
\ber
B_{1} &=& K_{X} - 2G_{\phi} + 2XG_{X\phi} - \frac{2X^{2}G_{X}^{2}}{\mpl^{2}}~,\nn\\
B_{2} &=& K_{XX} - 2G_{X\phi} + \frac{4 X^{2}G_{X}^{2}}{\mpl^{2}}~,\nn\\
B_{3} &=&  K_{X\phi} - G_{\phi\phi} + \l(\frac{G_{X}}{\mpl^{2}}\r)\l(4XG_{\phi} - XK_{X} - K\r)~,\label{Aeqn: B}\\
B_{4} &=& (\Box\phi)^{2} - (\partial^{\mu}\phi)_{\,;\,\nu}(\partial^{\nu}\phi)_{\,;\,\mu}~,\nn\\
B_{5} &=& (\partial^{\mu}\phi\,\partial_{\mu} X)\Box\phi - \partial^{\mu}X\,\partial_{\mu}X~.\nn
\eer
Eq.~(\ref{Aeqn: G_I EOM}) is valid in any curved space-time.
In a spatially flat FRW line element~(\ref{eqn: FRW}), it is straightforward to verify that Eq.~(\ref{Aeqn: G_I EOM}) reduces to Eq.~(\ref{eqn: KgEqn for G}).

Let us consider the following perturbation in the scalar field
\beq
\phi(\vec{x},t)\, = \,\phi(t)\,+\,\delta\phi(\vec{x},t)~\,
\label{Aeqn: phi + delta phi}
\eeq
where $\phi(t)$ is the solution of the background equation~(\ref{eqn: KgEqn for G}). Since scalar field perturbations propagates as sound waves only at scales much below the size of the horizon,  as a reasonable approximation for the derivation of $c_{_s}$, one can ignore the metric perturbations and consider the evolution of $\phi(\vec{x},t)$ defined in Eq.~(\ref{Aeqn: phi + delta phi}) in a spatially flat FRW line element~(\ref{eqn: FRW}).
On substituting Eq.~(\ref{Aeqn: phi + delta phi}) in Eq.~(\ref{Aeqn: G_I EOM}) and eliminating the background equation~(\ref{eqn: KgEqn for G}), we obtain the following equation of motion for $\delta\phi(\vec{x},t)$:
\beq
N_{1}\,\ddot{\delta \phi}\, -\, N_{2}\l(\frac{\nabla^{2}\delta\phi}{a^{2}}\r)\,+\,
N_{3}\,\dot{\delta\phi}\,+\,N_{4}\,\delta\phi \,=\,0~,
\label{Aeqn: G_I delta phi}
\eeq
where
\ber
N_{1} &=& K_{X} + 2XK_{XX} -  2G_{\phi}  - 2XG_{X\phi} + 6H\dot{\phi}\l(G_{X} + X G_{XX}\r) + \f{6X^{2}G_{X}^{2}}{\mpl^{2}}~,\nn\\
N_{2} &=& K_{X} - 2G_{\phi}  + 2XG_{X\phi} + 2\ddot{\phi}\l(G_{X} + X G_{XX}\r) + 4H\dot{\phi}G_{X} - \f{2X^{2}G_{X}^{2}}{\mpl^{2}}~,\nn\\
N_{3} &=& 6 H G_{X}\l(\ddot{\phi} + 2H \dot{\phi}\r) + G_{XX}\l(12 H^{2}X \dot{\phi} + 21 X \dot{\phi} \ddot{\phi}\r) + 3 H G_{XXX}\dot{\phi}^{4} \ddot{\phi} - K_{X\phi}\dot{\phi}\label{Aeqn: N}\\
&& ~ + ~3H B_{1} + 2 \dot{\phi} \ddot{\phi} B_{2} + 2\dot{\phi}B_{3}
+ (\ddot{\phi} + 3 H \dot{\phi})\dot{\phi}B_{1X} + 2 X \dot{\phi}^{2} B_{2X}  + 2 X \dot{\phi}B_{3X}~,\nn\\
N_{4} &=& 6 H \dot{\phi} G_{X\phi}\l(\ddot{\phi} + H\dot{\phi}\r) + 3H \dot{\phi}^{3}\ddot{\phi}G_{XX\phi} - K_{\phi\phi} + (\ddot{\phi} + 3 H \dot{\phi})B_{1\phi} + 2 X \dot{\phi} B_{2\phi}  + 2 X B_{3\phi}~.\nn
\eer
In the above equation,  $B_{1}$, $B_{2}$ and $B_{3}$ are defined in Eq.~(\ref{Aeqn: B}) and the notations such as $B_{1X}$ denotes $\partial B_{1}/\partial X$.
In the Fourier space $\nabla^{2}\delta\phi = -k^{2}\delta \phi_{_k}$. Therefore, at scales much below horizon, which corresponds the large $k$ limit, Eq.~(\ref{Aeqn: G_I delta phi}) approximately becomes a sound wave equation:
\beq
\ddot{\delta \phi}\, -\, c_{_s}^{2}\l(\frac{\nabla^{2}\delta\phi}{a^{2}}\r)\,\simeq\,0
\eeq
where $c_{_s}^{2} = N_{2}/N_{1}$ and therefore given by
\beq
c_{_s}^{2} = \frac{K_{X} - 2G_{\phi}  + 2XG_{X\phi} + 2\ddot{\phi}\l(G_{X} + X G_{XX}\r) + 4H\dot{\phi}G_{X} - 2X^{2}G_{X}^{2}/\mpl^{2}}
{K_{X} + 2XK_{XX} -  2G_{\phi}  - 2XG_{X\phi} + 6H\dot{\phi}\l(G_{X} + X G_{XX}\r) + 6X^{2}G_{X}^{2}/\mpl^{2}}~.
\label{Aeqn: sound speed}
\eeq
For the k-inflation models which corresponds to setting $G(X,\phi) = 0$ in the Lagrangian~(\ref{eqn: G-Lagrangian}), the above equation for $c_{_s}^{2}$ reduces to  Eq.~(\ref{eqn: sound speed k-inf def}). In addition, if $K(X,\phi) = X - V(\phi)$, which represents canonical scalar field models one gets back the standard result $c_{_s}^{2} = 1$~\cite{Hu-1998}.


\end{document}